\documentstyle[twoside,fleqn,espcrc2,epsfig]{article}

\def\aprge{\buildrel > \over {_{\sim}}}

\title{Underground Muon Physics with the MACRO experiment}

\author{M. Sioli\address{Dipartimento di Fisica dell'Universit\`a
di Bologna and INFN, I-40126 Bologna, Italy}
for the MACRO Collaboration\thanks{for a complete list of the Collaboration
see the contribution of L. Patrizii in these proceedings}}

\begin{document}

\begin{abstract}
Underground muon events detected by the MACRO experiment at Gran Sasso 
have been studied for different purposes.
The studies include the vertical muon intensity measurement, multiplicity 
distribution, lateral and angular muon distribution and searches for 
substructures inside muon bundles.
These analyses have contributed to bring new insights in cosmic ray physics, 
in particular in the framework of primary cosmic ray composition studies.
Moreover, this activity allows the testing and tuning of Monte Carlo
simulations, in particular for aspects associated with models of hadronic 
interactions and muon propagation through the rock.
\end{abstract}

\maketitle

\section{INTRODUCTION}
Muons detected deep underground are a useful tool for different 
physics and astrophysics items.
These muons are the decay product of mesons originating in the very 
first hadronic interactions in the atmosphere or in secondary
interactions during the shower development. Therefore, their study 
can provide many informations about primary cosmic rays (CR) 
composition and/or high energy hadronic interactions.
Muons arriving in the underground Gran Sasso Laboratory have crossed 
at least $h\ge$ 3100 $hg/cm^{2}$ of standard rock, corresponding to 
a muon energy cut $E_{\mu}\ge 1.3$ TeV at the surface. 
This means that the primary CR energies range from some TeV/nucleon 
up to the maximum energies well beyond the ``knee''.

The MACRO experiment has collected a large amount of muon data in the 
last decade, at a rate of $\sim 6.6 \times 10^{6}$ muon events/live year, 
and the $\sim$ 6\% of these are multiple muon events.
Many underground observables have been studied. The multiplicity
distribution, i.e. the rate of muon events as a function of their
multiplicity, is a quantity strongly dependent on the primary composition
model. A detailed analysis on primary composition has been performed 
by the MACRO collaboration \cite{macro_compo1,macro_compo2}:
one of the result is that data prefers a composition model with an average
mass slightly increasing with energy above the knee\footnote{for a more 
detailed discussion on composition studies see the contribution of 
E. Scapparone in these proceedings}.

Nevertheless, in the context of composition studies the knowledge of the 
hadronic interaction model is crucial. The main contribution to the 
systematic uncertainties in these analyses are due to the interaction
models adopted and it is important to find out new observables to
test the reliability of these models implemented in the Monte Carlo 
simulations. Moreover, is intrinsically important to test interaction 
models in kinematical regions not yet explored at accelerators or colliders.
For instance, about 5\% of MACRO muon data comes from $pp$ interactions 
with $\sqrt{s} >$ 2 TeV, while about 30\% of the muons observed are
the decay products of mesons produced with a (pseudo)rapidity $\eta_{cm} >$ 5.
The situation is more evident if we consider that part of primary 
interactions in the atmosphere are nucleus-nucleus interactions,
and very few data for energy $E_{lab}\aprge$ 150 A GeV are available.  
In the following, we will focus on the decoherence analysis and on
cluster analysis, two different tools able to extract informations
on the interaction model adopted in the simulation codes.

\section{THE MACRO DETECTOR}
The MACRO detector \cite{macro_detect} is a large area apparatus
located in hall B of the Gran Sasso Laboratory at an average depth
of 3800 hg/cm$^2$ of standard rock ($E_\mu\geq~$1.3 TeV).
It has a modular structure, organized in six almost identical 
``supermodules'' covering an horizontal surface area of $\sim 1000$~m$^2$.
The apparatus is equipped with three different and independent sub-detectors:
streamer tube chambers for particle tracking, scintillator counters
for timing and energy loss reconstruction and nuclear track-etch detectors
optimized for the search of magnetic monopoles.
The wire view of the streamer tube system is complemented with 
a second view, disposed at $26.5^\circ$ with respect to the wire view,
realized with aluminium pick-up strips. 
The spatial resolutions are $\sigma_{W}$=1.1~cm and $\sigma_{S}$=1.6~cm
for the wire and strip view respectively.
This arrangement allows the 3-D track reconstruction with an intrinsic
pointing resolution of $0.2^\circ$.

\section{SIMULATION TOOLS}
The full simulation chain used to interpret MACRO data is
composed of an event generator modelling high energy hadronic 
interactions, included in a shower propagation code which follows 
particles above threshold up to a given atmospheric depth. 
A muon transport code propagate muons inside the mountain
overburden the apparatus and a detector simulator produces an output
in the same format of real data.

\vspace{0.3cm}
\par$\bullet$ {\bf Event generator}

Most of the analyses of MACRO data have been carried on using the 
original HEMAS interaction model \cite{hemas}, based on the parametrizations 
of minimum bias events collected at the $Sp\overline{p}S$ collider at 
CERN \cite{ua5}. According to these results, the charged multiplicity 
is sampled from a negative binomial distribution and the transverse 
momentum contains a power law component. The model includes nuclear 
target effects and extrapolations to higher energies are performed 
in the context of log$(s)$ physics.

The interaction model in Ref. \cite{nim85} (called ``NIM85'') 
has been tested in a parametrized form with MACRO muon data
 \cite{old_deco,cluster}.
This model neglects some important experimental results: for instance,
the charged multiplicity is sampled from a Poisson distribution and
the transverse momentum distribution contains only a pure exponential 
functional form.

Presently new interaction models are under study: DPMJET \cite{dpmjet}, 
QGSJET \cite{qgsjet}, SIBYLL \cite{sibyll} and HDPM, the original interaction
model of the shower simulation code CORSIKA \cite{corsika}.
These are phenomenological models where the interactions are treated
at parton level, with the exception of the HDPM generator, which is based 
on the DPM model but it is built according to parametrized results.
They have the common feature to refer to the Regge-Gribov theories for 
the modelling of the soft part of the interactions, where perturbative 
QCD cannot be applied. 
Nevertheless, the transverse component of the interactions is not 
constrained by the theory and is introduced ``by hand'',
according to some experimental results such as the seagull 
effect \cite{seagull} or the Cronin effect \cite{cronin} in nuclear 
interactions.
In this context, it is useful the comparison of the model one
to each other to estimate the systematic uncertainty associated to
the unknown transverse structure of the interactions. 

\vspace{0.3cm}
\par$\bullet$ {\bf Cascade code}

Two different shower propagation codes have been used in the analyses:
HEMAS\footnote{the name HEMAS refers both to the hadronic interaction 
model and to the cascade code} \cite{hemas,hemasdpm} and 
CORSIKA \cite{corsika}.
HEMAS is conceived as a fast tool to compute the hadronic, electromagnetic 
and muon component of air showers.
It has been extensively used in the MACRO collaboration and has revised
many improvements since its first release.
It introduces some approximations in the shower development, so that the 
model can be used to follow only particles with a minimum energy in 
atmosphere $E>500$ GeV. The electromagnetic size is computed by means
a semi-analytical method.
It can be used interfaced with the DPMJET and HEMAS interaction models.

Recently, the CORSIKA \cite{corsika} Monte Carlo code, generally used
in surface EAS-arrays, has been interfaced with the muon transport code
to propagate muons up to the Gran Sasso depth. This code has been used
only in the analysis of high multiplicity events.

\newpage

\vspace{0.3cm}
\par$\bullet$ {\bf Muon transport code}

The muon propagation in the rock has been realized using
the PROPMU package \cite{propmu}, which represents an improvement
with respect to the propagation model included in the original 
HEMAS version.
It takes into account muon energy loss due to multiple Coulomb scattering
and to discrete processes, such as bremsstrahlung, pair production processes
and photonuclear interactions.

\vspace{0.3cm}
\par$\bullet$ {\bf Detector simulator}

The response of the apparatus is simulated by means a GEANT \cite{geant}
based code, called GMACRO. It reproduces all relevant physical processes 
in the detector and produces an output in the same format of real data, 
so as to process data using the same offline chain of real data.

\section{ANALYSIS RESULTS}

\subsection{Decoherence Function}

\begin{figure}[thb]
\epsfig{file=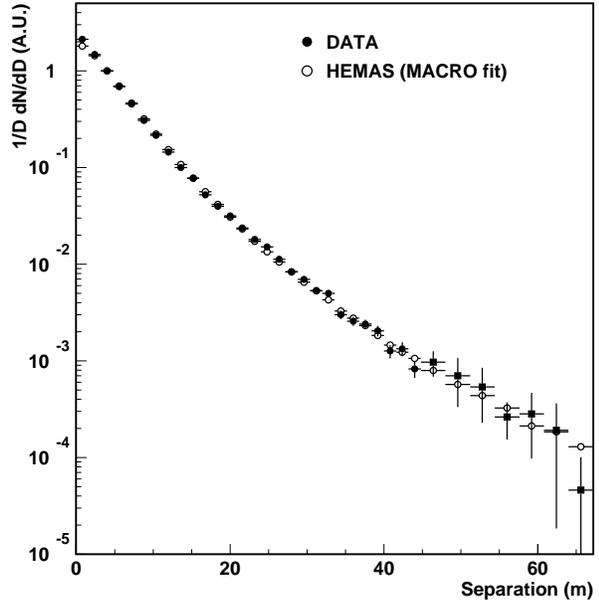,width=8cm}
\caption{Experimental decoherence distribution (corrected for the
detector effects) compared with the predictions of the HEMAS Monte Carlo
(MACRO-fit primary composition model)}
\label{f:decounf}
\end{figure}

The {\it decoherence function}, defined as the distribution of the
muon pair separation in multiple muon events, is mainly connected
with the muon lateral distribution with respect to the shower axis
and is very sensitive to the transverse structure of the 
hadronic interaction models.
The decoherence distribution is instead weakly dependent on the 
primary composition model, as far as the shape of the distribution 
is concerned. Therefore, the study of this function allows to 
some extent the disentangle between the two effects.
MACRO has studied the decoherence function up to the maximum distance
allowed by the apparatus \cite{new_deco} ($\sim $ 70 m). 
The unfolded decoherence function (the distribution corrected for the 
detector effects) is shown in Fig.\ref{f:decounf}, where it is compared 
with the predictions of the HEMAS Monte Carlo.

A further check on the reliability of the simulation code has been 
performed in different rock/zenith windows, constraining some component 
of the shower development with respect to others \cite{new_deco}.
Again, the comparison shows a good agreement between data and Monte Carlo.

\begin{figure}[thb]
\epsfig{file=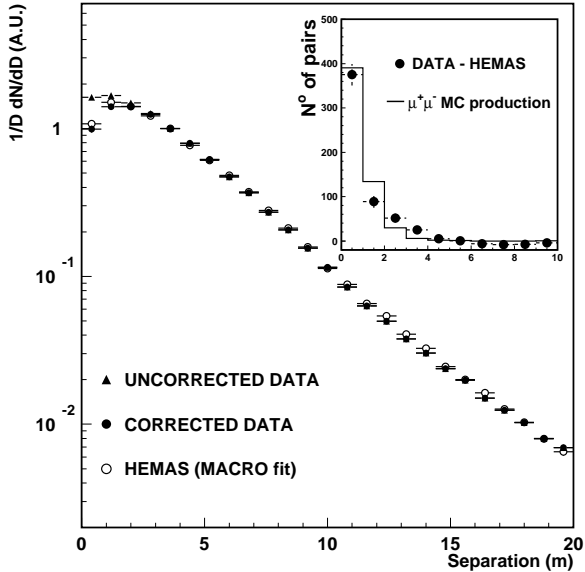,width=8cm}
\caption{Zoom of the low distance region of the decoherence function
before and after the inclusion of muon pair production process, and
comparison with the Monte Carlo. In the inset it is shown the difference
of the distributions before and after the correction, compared with
the expectations of the Monte Carlo.}
\label{f:decomumu}
\end{figure}

Finally, the study of the decoherence function in the 
very low distance region has shown that the QED process 
$\mu^\pm +N \rightarrow \mu^\pm + N+\mu^+ + \mu^-$ 
at small distances must be taken into account if we want 
to reproduce the experimental data \cite{isvhecri}.
This is shown in Fig.\ref{f:decomumu}, where the low distance
region of the decoherence function is shown before and after the
correction. The contribution of this process is negligible compared 
to the $e^{+}e^{-}$ pair production process in the GeV range, but it 
becomes progressively more important in the TeV region \cite{olga}.

\subsection{Cluster Analysis}
The search for substructures inside muon bundles is able to provide 
additional information with respect to traditional methods \cite{cluster}.
In some events muon bundles appear to be splitted into ``clusters''
and we ask if this feature is the result of simple statistical fluctuations 
in the muon lateral distribution or if there is some dynamical correlation 
connected with the development of the shower in the atmosphere.
From an experimental point of view, we select muon bundles with at least 
8 muons underground ($N_{\mu} \ge$ 8) corresponding to CR primary energies
$E_{primary}\aprge $ 1000 TeV. The search for muon clusters is performed
by means of an iterative cluster finding algorithm, which groups the muons 
depending on the choice of a free parameter called $\chi_{cut}$
(for the definition of this parameter see Ref. \cite{cluster}).

\begin{figure}[b!]
\epsfig{file=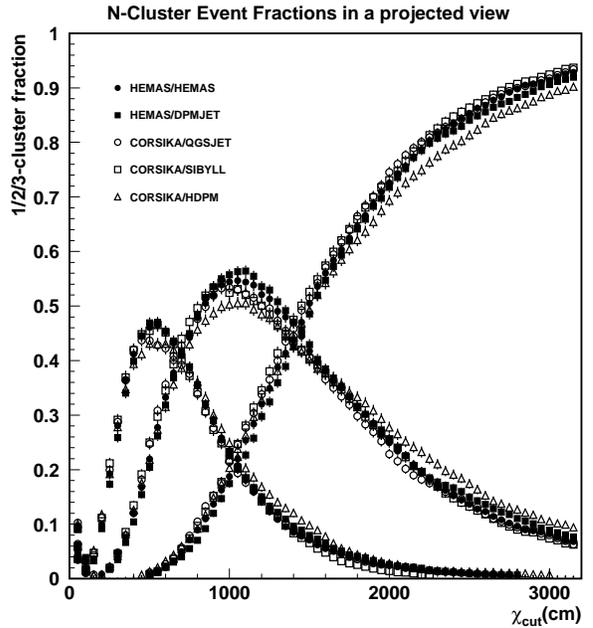,width=8cm}
\caption{Fraction of events with 1,2 and 3 clusters in muon bundles
for different hadronic interaction models. The detector effects have
been considered and all the simulations have been obtained with the
MACRO-fit composition model.}
\label{f:cluster}
\end{figure}

In Ref. \cite{cluster} has been pointed out that this method is sensitive 
both on the primary composition model and on the hadronic composition model.
The comparison was made between two extreme composition model 
(the ``heavy'' and ``light'' composition models \cite{adelaide})
and between two very different hadronic interaction model (HEMAS and NIM85).
Most of the effect can be explained as the result of fluctuations of the
muon {\it density} inside the bundles. Considering that the density 
(namely the average number of muons per unity of area) depends both on 
primary mass number and on the modelling of meson transverse momentum, 
we can expect that the cluster effect is sensitive to the composition 
model and to the hadronic interaction model at the same time.
On the other hand, if we consider the interaction models quoted in previous 
sections, the sensitivity of the cluster effect on the interaction model 
becomes weaker. This is shown in Fig.\ref{f:cluster}, where is reported 
the cluster rates as a function of the parameter $\chi_{cut}$ for different
interaction models and for a fixed primary composition model (the model 
derived from the fit of the MACRO multiplicity distribution 
 \cite{macro_compo2}).
In this case we are forced to enhance the sensitivity applying some 
selection criteria which correlate the underground substructures with 
the hadronic interaction features in the atmosphere.

\begin{figure}[b!]
\epsfig{file=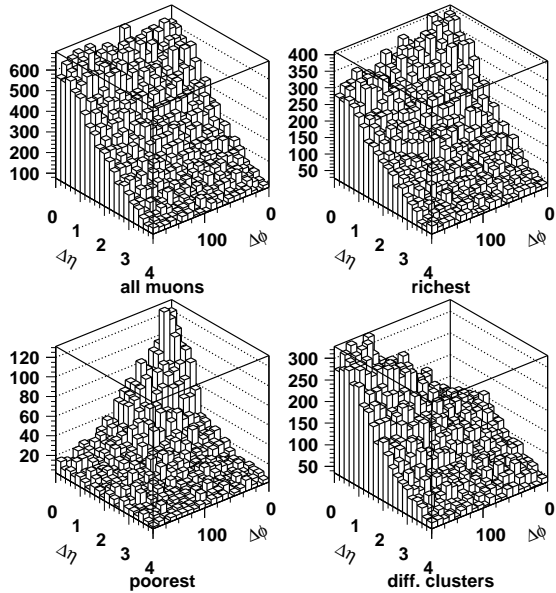,width=8cm}
\caption{2-dimensional distributions of the distance of first interaction 
mesons in the $\eta-\phi$ space. It is shown the case of no selection,
of the selection of muons belonging to the same cluster (poorest or richest)
or to different clusters.}
\label{f:lego}
\end{figure}

Apart from statistical fluctuations, a Monte Carlo study has revealed 
other two mechanisms responsible for the cluster effect:
\par$\bullet$ muons belonging to the same cluster have a larger probability
to have a common parent meson in the steps of shower tree generation;
\par$\bullet$ muons belonging to the same cluster are the decay products of
mesons highly correlated in the phase space of the very first hadronic
interactions in the atmosphere.
We considered only events reconstructed as two-cluster events by
the algorithm with a fixed $\chi_{cut}$ and we studied the kinematical 
variables pseudorapidity $\eta$ and azimuthal angle $\phi$ of the parent 
mesons in the first interaction of the shower. The typical topology
of these selected events is a muon rich cluster close to the shower axis,
generally the remnant of the shower development after many steps, and a 
far cluster with few muons directly generated in the very first 
hadronic interactions.
Fig.\ref{f:lego} shows the distributions of the relative distance in the 
$\eta-\phi$ space of pairs of muon parent mesons originated in the first 
interactions: the topological selection of muons belonging to the same or 
to different clusters reflects in a strong selection of different phase 
space regions of hadronic interactions. 

A quantitative computation of the relative contributions of these effects 
is at presently under study.

\section{CONCLUSIONS}
The MACRO experiment has collected a large amount of muon data in the
last decade. The analyses performed with these data allowed to draw 
conclusions about several items of muon physics.
The dimension and granularity of the detector allowed the detection of 
multiplicity up to $\sim 40$, under more than 3000 hg/cm$^2$ of rock 
overburden. The study of the multiplicity distribution has been used 
to extract informations on the primary cosmic ray composition. 

The modelling of the transverse component of the interaction models at 
TeV energies, connected with the lateral distribution of cosmic 
ray shower, is one of the main source of uncertainties in the simulation 
codes. The analysis of the decoherence function has shown the reliability 
of the HEMAS Monte Carlo, used in most of the MACRO analyses.

The study of second order effects, like the search for jet substructures 
inside muon bundles, is a useful tool to add new informations with 
respect to conventional analyses. 
The physical interpretation of the results has shown the dynamical 
origin of the effect, connected with the development of the shower 
in the atmosphere.

\end{document}